\documentstyle[epsf]{elsart} 
\begin{document}    

\def\be{\begin{equation}}
\def\ee{\end{equation}}
\def\bea{\vspace{-0.5in}\begin{eqnarray}}
\def\eea{\end{eqnarray}}
 
\parbox{1.5in}{ \leftline{WM-02-102}
                           \leftline{JLAB-THY-02-08 }}
\vspace{0.2in}

\begin{frontmatter}
\title{The role of interaction vertices in bound state calculations}
\author{\c{C}etin \c{S}avkl{\i}$^{1}$, Franz Gross$^{1,2}$, and John Tjon$^{3}$}
\address {$^1$Department of Physics, College of William and Mary,
Williamsburg, Virginia 23187, USA\\
$^2$Jefferson Lab,
12000 Jefferson Avenue, Newport News, VA 23606, USA\\
$^3$Institute for Theoretical Physics, University of Utrecht,
Princetonplein 
P.O. Box 80.006, 3508 TA Utrecht, the Netherlands}

\begin{abstract}
In recent studies of the one and two-body Greens' function
for scalar interactions it
was  shown that crossed ladder and ``crossed rainbow'' (for the one-body
case) exchanges play a crucial role in nonperturbative dynamics.  In this
letter we use exact analytical and  numerical results to show that
the contribution of vertex dressings to the two-body bound state mass for
scalar QED are cancelled by the self-energy and
wavefunction normalization.  This proves, for
the first time, that the mass of a two-body bound state given by the {\it
full\/} theory  can in a very good approximation
be obtained by summing only ladder and crossed ladder
diagrams using a bare vertex and a {\it constant\/} dressed mass.  We also
discuss the implications of the remarkable cancellation between rainbow and  
crossed rainbow diagrams that is a feature of one-body calculations.  
\end{abstract}




\date{\today} 
\end{frontmatter}

\section{Introduction}
\label{introduction}

In general, a proper description of bound states
requires an infinite summation of all possible interactions. Since this is
usually  not an easy task various approximation methods are used. Perhaps
the best  known approximations for the one-body and two-body systems are,
respectively,  the rainbow and the ladder approximations.  The
Dyson-Schwinger equation is usually used to sum the rainbow diagrams for the
one-body propagator, and the Bethe-Salpeter equation {\it in ladder
approximation\/} sums the two-body ladders exactly. In both cases these
kernels do not sum any crossed exchanges.  We now know that, for scalar
theories, crossed ladder exchanges make a very significant contribution to
the binding energies of two-body systems~\cite{TACO,TACO2}, and that
``crossed rainbow'' diagrams make equally significant contributions to the
one-body dressed mass~\cite{SAVKLI1}.  This means that equations that
include crossed exchanges approximately do a better job of reproducing the
exact result than does the Bethe-Salpeter equation (in ladder
approximation) or the Dyson-Schwinger equation (in rainbow approximation). In
this letter our goals are (i) to study the role of vertex corrections in the
two-body Greens' function, and to (ii) to discuss
the remarkable implication of the cancellation of rainbow and crossed
rainbow diagrams in the one-body propagator.  Here, and in the next
section, we only discuss vertex corrections to the two-body Greens' function, and
return to the one-body propagator in Sec.~III below.  

\begin{figure}
\begin{center}
\mbox{
    \epsfxsize=3.2in
\epsffile{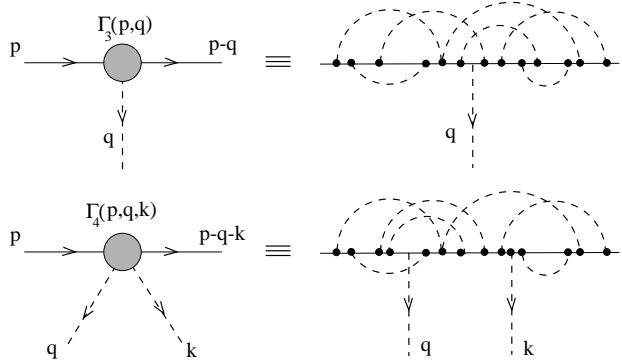}
}
\end{center}
\caption{Because of the interactions, the one-particle
irreducable vertex functions $\Gamma_n$ $(n=3,4,\cdots)$ depend on the
external momenta. }
\label{definition}
\end{figure}

\begin{figure}
\begin{center}
\mbox{
    \epsfxsize=2.3in
\epsffile{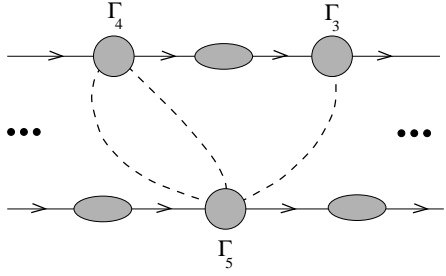}
}
\end{center}
\caption{ Exact computation of the two-body bound state propagator requires
the summation of all particle self energies, vertex corrections, and ladder
and crossed ladder exchanges.}
\label{2body}
\end{figure}

In general a consistent treatment of any nonperturbative calculation 
must involve summation of all possible vertex corrections. Vertex corrections
are those irreducible diagrams that surround an interaction vertex. Take
$\phi^3$ theory as an example. The elementary vertex is the
three-point  vertex, $\Gamma_3$, but the particle interactions will lead to
the appearance of $n$th order irreducable vertices, $\Gamma_n$, as
illustrated in Fig.~\ref{definition}.
The propagation of a bound state therefore involves a summation of all
diagrams with the inclusion of higher order vertices (Fig.~\ref{2body}). A
rigorous  determination of all of these vertices is in general not feasible. In the
literature on  bound states $\Gamma_{n>3}$ interaction vertices are usually
completely  ignored. The 3-point vertex $\Gamma_{3}$ can be approximately
calculated in  the ladder approximation~\cite{MARRIS}. However a rigorous
determination of  the {\em exact} form of the 3-point vertex is difficult,
for this requires  the knowledge of even higher order vertices.
 
In order to be able to make a connection between the exact theory and
predictions based on approximate bound  state equations it is essential that
the role of interaction vertices be understood. The Feynman-Schwinger
Representation (FSR)  is a useful technique for this purpose. The FSR is an
approach based on  Euclidean path integrals similar to lattice gauge theory
\cite{TACO,TACO2,SAVKLI1,TACO3,FSR,SIMONOV1,SIMONOV2,SIMONOV3,BRAMBI,SAVKLI2,SAVKLI3}. 
In  this approach the path integrals over quantum  fields
are integrated out and replaced by path integrals over  the trajectories of
the particles. 

In this paper we determine exactly the bound state mass for two scalar particles in
scalar QED in the quenched approximation, i.e. neglecting only charge loop contributions,
but including all self-energy, vertex and crossed ladder contributions.
We in particular demonstrate, for the first time, that the full bound state
result dictated by a Lagrangian can be well approximated by summing only generalized
ladder diagrams (``generalized'' ladders include crossed ladders and, in
theories with an elementary four-point interaction, both overlapping and
non-overlapping ``triangle'' and ``bubble'' diagrams). In the next section
we investigate the interplay of vertex, self-energy and wave function normalizations
within the context of massive scalar QED (SQED). First we look at the
implications in 0+1  dimensions, where exact analytic results can be
obtained using the FSR  method. Next we extend the analysis to 3+1
dimensions using numerical methods, and show that in SQED$_{3+1}$ wave
function normalization and vertex function normalization exactly cancel.

\section{Determination of vertex contributions using the FSR approach }

The Minkowski
metric expression for the SQED Lagrangian in Feynman gauge is 
\bea
{\cal{L}}_{\rm SQED}&=&-m^2\chi^2-\frac{1}{4}F_{\mu\nu}^2+\frac{1}{2}\mu^2A_\mu^2-
\frac{1}{2}(\partial^\mu A_\mu)^2\nonumber\\
&+&(\partial_\mu-ieA_\mu)\chi^*(\partial^\mu+ieA^\mu)\chi\, , 
\label{lag}
\eea
where $A_\mu$ is the gauge field of mass $\mu$, $\chi$ is a charged field of
mass $m$ and charge $e$,  $F_{\mu\nu}=\partial_\mu A_\nu-\partial_\nu
A_\mu$,  and
$A_\mu^2=A_\mu A^\mu$, $F_{\mu\nu}^2=F_{\mu\nu}F^{\mu\nu}$.  

The final FSR result for the  two-body propagator involves a quantum
mechanical path integral that sums up  contributions coming from all
possible {\em trajectories} of the two {\em charged particles} $\chi$.  This
path integral is
\be
G\simeq \bigg\lmoustache_0^\infty 
ds\,\, 
\bigg\lmoustache_0^\infty 
d\bar{s}\,\, \bigg\lmoustache\, 
({\cal D}z)_{xy}\,
\bigg\lmoustache\, 
({\cal D}\bar{z})_{\bar{x}\bar{y}}
e^{ -K[z,s]-K[\bar{z},\bar{s}] }    
\langle W(C)\rangle\, ,
\label{gfin}
\ee
where the particle trajectories $z_i(\tau)$ and $\bar{z}_i(\tau)$ are
parametric functions of the parameter $\tau$, with endpoints $z_i(0)=x_i$,
$\bar{z}_i(0)=\bar{x}_i$, $z_i(1)=y_i$, and
$\bar{z}_i(1)=\bar{y}_i$, with $i=$1 to 4.  The kinetic term is defined by
\bea
K[z,s]&=&m^2s+\frac{1}{4s}\int_0^1 d\tau \,\dot{z}^2(\tau),
\eea
and the Wilson loop average $\langle W(C)\rangle$, obtained in this case by
an analytic integration over the fields $A_\mu$, is 
\bea
\langle W(C)\rangle&=&{\rm exp}\,\biggl[-\frac{e^2}{2}\int_C\,dz_\mu\,\int_C
\,d\bar{z}_\nu
\,\Delta_{\mu\nu}(z-\bar{z},\mu)\biggl],\label{v.sqed}\nonumber\\
\Delta_{\mu\nu}(x,\mu)&=& g_{\mu\nu}\bigg\lmoustache \frac{d^4p}{(2\pi)^4}
\frac{e^{ip x}}{p^2+\mu^2}\, ,\nonumber
\eea
where the contour $C$ goes from $x\to y \to \bar{y}\to \bar{x}\to x$.
In the numerical calculations the ultraviolet singularities are regulated 
by using a double Pauli-Villars subtraction
\bea
\Delta_{\mu\nu}(x,\mu)&=& g_{\mu\nu}\bigg\lmoustache \frac{d^4p}{(2\pi)^4}
\frac{e^{ip x}
(\Lambda_1^2-\mu^2)(\Lambda_2^2-\mu^2)}{(p^2+\mu^2)(p^2+\Lambda_1^2)(p^2+
\Lambda_2^2)}\, .\nonumber
\eea
In the limit of large $z_4(1)=\bar{z}_4(1)=T$, the ground state mass is given
by
\begin{equation}
M_2=\lim_{T\rightarrow\infty}-\frac{d}{dT} {\rm ln}[G(T)]= \frac{\int {\cal
D}Z S'[Z]e^{-S[Z]}}{\int {\cal D}Z e^{-S[Z]}}\, .
\label{groundstate}
\end{equation}

Equation~(\ref{gfin}) has a very nice physical interpretation.  The
term $\Delta_{\mu\nu}(z_a-z_b,\mu)$ describes the propagation of gauge field
interations between any two points on the particle trajectories, and the
appearance of these interaction terms in the exponent means that the
interactions are summed to all orders with arbitrary ordering of the points
on the trajectories.  Self-interactions come from terms with the two points
$z_a$ and $z_b$ on the {\it same\/} trajectory, generalized ladder
exchanges arise if the two points are on {\it different\/} trajectories, and
vertex corrections arise from a combination of the two.

\subsection{SQED$_{0+1}$}

In order to understand the role of vertex corrections we first consider the
simple case of SQED in 0+1 dimension. This interaction has been 
discussed in detail in Refs~\cite{TACO2,SAVKLI1,SIMONOV3,SAVKLI3}. 

In 0+1 dimension the FSR formulation yields the exact, {\it analytic\/}
result for the $n$-body bound state interaction energy.  It is
\bea
V_{\rm total} &=& V_{\rm self}+V_{\rm exchange}\nonumber\\
 &=&n\,\frac{e^2}{2\mu^2}-n\,(n-1)\,\frac{e^2}{2\mu^2}\, ,
\label{sqed1}\\
&=&-n\,(n-2)\,\frac{e^2}{2\mu^2}\, ,
\label{sqed.fsr.mass}
\end{eqnarray}
where the first term in (\ref{sqed1}) is from self-energy corrections, and
the second from exchange contributions. The exact
$n$-body dressed mass for SQED in 0+1 dimension is
then~\cite{TACO2,SAVKLI1,SAVKLI3}
\bea
M_n &=&n\,m + V_{\rm total}\, ,\nonumber\\
    &=&n\,m-n\,(n-2)\,\frac{e^2}{2\mu^2}\, ,\\
    &=&n\,M_1 + V_{\rm exchange}\, ,\label{sqed2} 
\end{eqnarray}
We emphasize that vertex corrections in the FSR
approach are automatically taken into account by combinations of self
energy and exchange interactions, and because these two interactions are
{\it additive\/} in 0+1 dimension, the vertex corrections are identically
zero!  This fact is summarized in Eq.~(\ref{sqed2}), which shows that the
{\it exact\/} result for the $n$-body bound state can be written as the
sum of $n$ {\it dressed\/} single particle masses plus energy arising from
the generalized ladder exchanges (only).   

\begin{figure}
\begin{center}
\mbox{
    \epsfxsize=3.3in
\epsffile{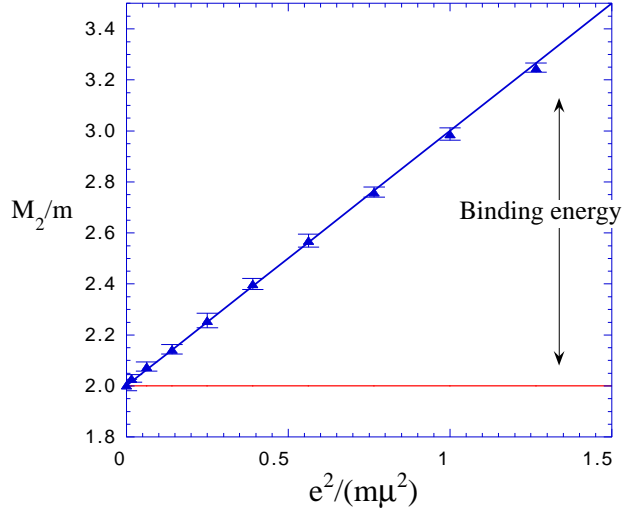}
}
\end{center}
\caption{The two-body bound state mass for SQED in 0+1 dimension.  The upper line is
the theoretical result for 2$M_1$, with numerical results that agree with the analytic
results, an important check.  The lower line with $M_2/m$=2 is the analytical result
for $M_2^{\rm exch}=M_2^{\rm tot}$.  (Compare this with Fig.~\ref{mvsg2.dressed.eps}.) }
\label{0+1}
\end{figure}

In Fig.~\ref{0+1} we display the one-body and two-body bound state results. 
The figure shows the effects of higher
order interaction vertices are included if one uses the
{\it dressed\/} mass obtained from the original Lagrangian, the {\it
bare\/} vertices with the original coupling strength $e$, and then sums all
exchange interactions. This statement may be symbolically expressed by
\bea
L(e,\mu,m)\ +\ V_{\rm total} &\longrightarrow& L(e,\mu,M_1) + V_{\rm exchange}
\end{eqnarray}
where $V_{\rm total}$ implies that all interactions are summed.

It might appear that these results are an oddity of 0+1 dimension.  In the
next section we will show that they also hold for 3+1 dimensions.

\subsection{SQED$_{3+1}$}
\begin{figure}
\begin{center}
\mbox{
    \epsfxsize=3.4in
\epsffile{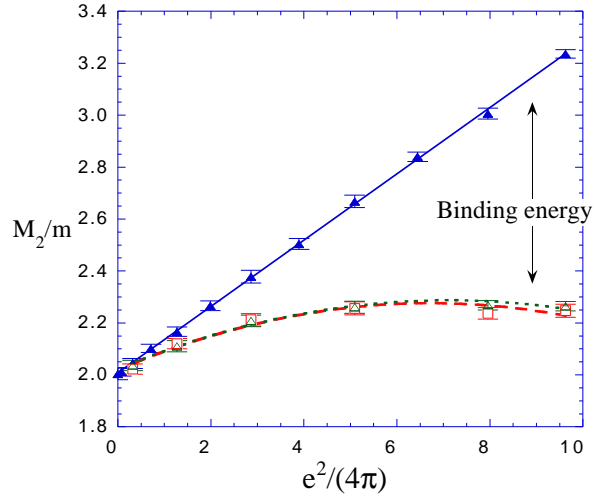}
}
\end{center}
\caption{ Two-body bound state mass for SQED in 3+1 dimensions. Solid
triangles are 2$M_1$, open squares are $M_2^{\rm exch}$, and open triangles
are $M_2^{\rm tot}$.  Here $\mu/m=0.15$, $\Lambda_1/\mu=3$, and
$\Lambda_2/\mu=5$.  The smooth lines are fits to the ``data''.  Note that 
$M_2^{\rm exch}=M_2^{\rm tot}$ to within errors.  (Compare this with
Fig.~\ref{0+1}.) }
\label{mvsg2.dressed.eps}
\end{figure}

We adopt the following procedure for determining the contribution of vertex
corrections in 3+1 dimension. We start  with an initial bare mass $m$ and
calculate the full two-body bound state result with the inclusion of all
interactions: generalized ladders, self energies and vertex corrections.
Let us denote the result for the exact two-body bound state mass by $M_2^{\rm
tot}(e^2,m)$, since it will be a function of the coupling strength
$e$ and the bare input mass $m$, and the superscript ``tot'' implies that all
interactions are summed. Next we calculate the dressed one-body mass
$M_1(e^2,m)$. Then using the dressed mass value $M_1(e^2,m)$ we calculate the
bound state  mass $M_2^{\rm exch}(e^2,M_1)$ {\em by summing only the
generalized exchange interaction  contributions}. In this last calculation
we sum only exchange interactions (generalized ladders), but the
self energy is approximately taken into account since we use the (constant)
dressed one-body mass as input. However the vertex  corrections and
wavefunction renormalization are completely left out since we  use the
original vertex provided by the Lagrangian. In order to compare the full
result where all interactions have been summed with the result obtained by
two dressed particles interacting only by generalized ladder
exchanges we plot the bound state masses obtained by these methods.
Numerical results are presented in  Fig.~\ref{mvsg2.dressed.eps}.  This
result is qualitatively similar to that obtained analytically for SQED in 
0+1 dimension.   
%
%

\begin{figure}
\begin{center}
\mbox{
    \epsfxsize=3.2in
\epsffile{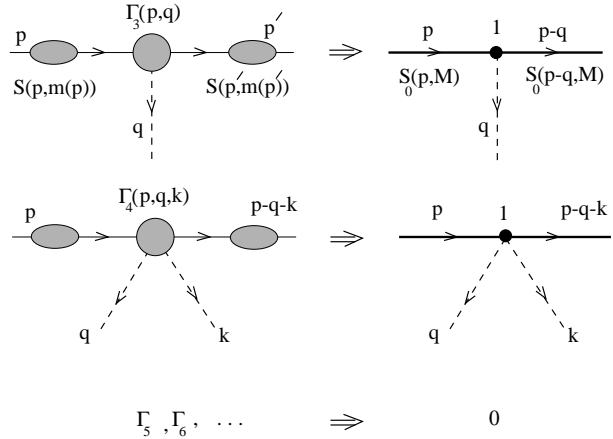}
}
\end{center}
\caption{ The correct two-body result can be obtained by simply using a
dressed constituent mass and a bare vertex, and ignoring the contributions
of higher order vertices.}
\label{mvsg2.sqed}
\end{figure}

To summarize the analytical and numerical results we have presented here
we give the following prescription for bound state calculations: 
In order to get the full result for bound states it is 
a good approximation to first solve for dressed one-body masses exactly
(summing all generalized rainbow diagrams), and then use these
dressed masses and the  bare interaction vertex provided by the Lagrangian
to calculate the bound  state mass by summing only generalized ladder
interactions (leaving out  vertex corrections). In terms of Feynman graphs
this prescription can  be expressed as in Fig.~\ref{mvsg2.sqed}

\section{Discussion}

The significance of the results presented above rests in the fact that the
problem of calculating exact results for bound state masses in SQED has been
reduced to that of calculating only generalized ladders. Summation of
generalized ladders can be addressed within the  context of bound state
equations~\cite{GROSS1,MWP}. Here, for the first time, we  have demonstrated
the connection between the full prediction of a Lagrangian and the summation
of generalized ladder diagrams.  Our results are rigorous for
SQED, but are only suggestive for more general theories with spin or internal
symmetries.  Since we have neglected charged particle loops (our
results are in quenched approximation), and the current is conserved in
SQED, it is perhaps not surprizing that the bare coupling is not
renormalized, but the fact that the momentum dependence of the dressed mass
and vertex corrections seem to cancel is surprizing and unexpected.  If we were
to unquench our calculation, or to use a  theory without a conserved current,
it is reasonable to expect that {\it both\/} the bare interaction and the
mass would be renormalized. 

Finally, we call attention to a remarkable cancellation that occurs in the
one-body calculations.  The exact self energies shown in Figs.~\ref{0+1}
and \ref{mvsg2.dressed.eps} are nearly linear in $e^2$
\cite{SAVKLI1}.  This remarkable fact implies that the {\it exact\/} self
energy is well approximated by the lowest order result from perturbation
theory.  It is instructive to see how this comes about.  If we expand the
self energy to fourth order, expanding each term about the bare mass $m$, we
have
\bea
S^{-1}_d(p^2) &=&m^2-p^2 + \Sigma(p^2) \nonumber\\
    &=&m^2-p^2 + \Sigma_2 +(p^2-m^2)\Sigma_2' +\Sigma_4\, ,\quad
\label{self1}
\end{eqnarray}
where $\Sigma_\ell=\Sigma_\ell(m^2)$ is the contribution of order $e^\ell$
evaluated at $p^2=m^2$, $\Sigma'=d\Sigma(p^2)/dp^2$ evaluated at $p^2=m^2$, 
and the formula is valid for $p^2-m^2\simeq e^2$.  Expanding the dressed
mass in a power series in $e^2$
\bea
M_1^2 &=&m^2 + m_2^2 +m_4^2 +\cdots \, , \label{mass1}
\end{eqnarray}
where $m_\ell^2$ is the contribution of order $e^\ell$, and substituting
into Eq.~(\ref{self1}), give
\bea
M_1^2 &=&m^2 + \Sigma_2 +\Sigma_2'\Sigma_2+\Sigma_4 +\cdots \, ,
\label{mass2}
\end{eqnarray}
The mass is then
\bea
M_1 &=&m + \frac{\Sigma_2}{2m}
+\frac{4m^2\left[\Sigma_2'\Sigma_2+\Sigma_4\right]-\Sigma_2^2}{8m^3} +\cdots
\, , \qquad \label{mass3} \\
&&\nonumber
\end{eqnarray}
The linearity of the exact result implies that the forth order term in
Eq.~(\ref{mass3}) must be zero (or very small), and this can be easily
confirmed by direct calculations!  

The cancellation of the fourth order mass correction (and all higher orders)
is reminisent of the cancellations between generalized ladders that explains
why quasipotential equations are more effective that the ladder
Bethe-Salpeter equation in explaining two-body binding energies.  It shows
that a simple evaluation of the second order self energy at the bare mass
point is more accurate than solution of the Dyson Schwinger equation in
rainbow approximation.

The general lesson seems to be that attempts to sum a small subclass of
diagrams exactly is often less accurate than the approximate summation of a
larger class of diagrams.

\section{Acknowledgement}
This work was supported in part by the US Department
of Energy under grant No.~DE-FG02-97ER41032.  The Southeastern 
Universities Research Association (SURA) operates the Thomas
Jefferson National Accelerator Facility under DOE contract
DE-AC05-84ER40150.


\end{document}